\documentclass[11pt]{article}

\date{}

\usepackage{url}
\usepackage{cite}
\usepackage{plain}

\usepackage{wrapfig}
\usepackage{graphics}
\usepackage{picins,graphicx}
\usepackage[english]{babel}
\usepackage[leqno]{amsmath}
\usepackage{amssymb}
\usepackage{verbatim}
\usepackage[mathscr]{eucal}

\usepackage{amstext}
\usepackage{makeidx}
\usepackage{amsthm}

\usepackage{hyperref}
\hypersetup{colorlinks,%
citecolor=red,%
linkcolor=blue,%
urlcolor=black}

\makeindex

\newtheorem{theorem}{Teorema} 

\newtheorem{proposition}{Proprieta'}
\newtheorem{definition}{Definizione}
\newtheorem{notation}{Nota}
\newtheorem{ex}{Esercizio} 
\newtheorem{esempio}{Esempio}

\newcommand{\vs}{\vspace{3mm}}
 
\newcommand{\no}{\noindent} 

\newcommand{\beq}{\begin{equation}} 
\newcommand{\eeq}{\end{equation}}

\newcommand{\bex}{\begin{ex}} 
\newcommand{\eex}{\end{ex}} 

\newcommand{\bese}{\begin{esempio}} 
\newcommand{\eese}{\end{esempio}} 

\newcommand{\bpro}{\begin{proposition}} 
\newcommand{\epro}{\end{proposition}} 

\newcommand{\ds}{\displaystyle}

\newcommand{\bthe}{\begin{theorem}} 
\newcommand{\ethe}{\end{theorem}}

\newcommand{\bnote}{\begin{notation}} 
\newcommand{\enote}{\end{notation}}

\newcommand{\bdefi}{\begin{definition}} 
\newcommand{\edefi}{\end{definition}} 

\newcommand{\bc}{\begin{center}} 
\newcommand{\ec}{\end{center}}

\newcommand{\mail}[1]{\href{unina:#1}{\texttt{#1}}}

\usepackage{pdfsync}

\author{Monica De Angelis\thanks{Univ. of Naples  "Federico II", Dip. Mat. Appl. "R.Caccioppoli", \newline
 Via Claudio n.21, 80125, Naples, Italy.
\newline\mail{modeange@unina.it}}}

\title{ A priori estimates for excitable models}

\begin{document}
\maketitle

\begin{abstract}
\vs \no The reaction-diffusion system of Fitzhugh Nagumo is considered. The initial-
boundary problems with Neumann and Dirichlet conditions are analyzed. By
means of an equivalent semilinear integrodifferential equation which characterizes
several dissipative models of viscoelasticity, biology, and superconductivity, some
results on existence, uniqueness and a priori estimates are deduced both in the linear
case and in the non linear one.

\vs \no {\bf{Keywords}}:{ Reaction - diffusion systems;\hspace{2mm} Biological applications;\hspace{2mm} Laplace  transform,\hspace{2mm}FitzHugh Nagumo model.}

\vs \no \textbf{Mathematics Subject Classification (2000)}\hspace{1mm}35E05  \hspace{1mm}35K35\hspace{1mm} 35K57 \hspace{1mm}35Q53 \hspace{1mm} 78A70
\end{abstract}


\vspace{3mm}


\section{Introduction }

\setcounter{equation}{0}
\setcounter{figure}{0}

\setcounter{definition}{0}
\setcounter{notation}{0}
\setcounter{theorem}{0}
\setcounter{esempio}{0}
\setcounter{ex}{0}
\setcounter{proposition}{0}
\setcounter{criterio}{0}

\vs As it is well known,the  FitzHugh - Nagumo model (FHN) has been introduced by FitzHugh (1961)  and Nagumo  et al.(1962) as a simplified description of the excitation and propagation of nerve impulses\cite{ks,i}.   
              
   Of course,  there are  many  other important physiological applications in addition to that for the propagation of nerve action potentials  modeled  by the  FHN system. One  such important application is related to the waves that arise in muscle tissue, particularly heart muscle and particularly  interesting is the biophysical phenomenon of reentry which occurs in the excitable cells of the heart. \cite{aa}
Another example is the reverberating cortical depression waves in the brain cortex.\cite{m2}
 As for the analysis of the (FHN) model,there exists a large bibliography both in the linear case and non linear problem.\cite{f,gt,rk,drw}. 

The  model is given by the set of p.d.e 
   \cite{m1,m2}: 

\beq     \label{12}
  \left \{
   \begin{array}{lll}
    \displaystyle{\frac{\partial \,u }{\partial \,t }} =\,  \varepsilon \,\frac{\partial^2 \,u }{\partial \,x^2 }
     \,-\, v\,\,  + f(u ) \,  \\
\\
\displaystyle{\frac{\partial \,v }{\partial \,t } }\, = \, b\, u\,
- \beta\, v\,.
\\

   \end{array}
  \right.
 \eeq

\no where  $\, u\,( \,x,t\,)$  models the transmembrane voltage  of a nerve axon at distance x and  time t, $\, v\,(\,x,t\,)$ is an auxiliary variable that  acts as recovery variable. Further, the  diffusion coefficient $\, \varepsilon \,$  and the parameters  $  \,  b, \, \beta \,$ are  all non negative  \cite {kss,f61}.

Besides, the  function $\,f (u)\,$  has the qualitative form of a cubic polynomial

\beq      \label{13}
f(u)\, =\,-\,a\,u\, +\,\varphi(u) \quad with \quad  \varphi \,=\, u^2\, (\,a+1\,-u\,)
\eeq

\vs \no where the parameter $ a $ is the threshold constant and generally one has $ \,0<a<1.\ $ However, in many papers  phenomena for the (FHN) system has been investigated considering $ a $ as a non homogeneous  function and $ a \,<\,0. $ \cite{d, gar}

 Moreover,    
\vs   Denoting by  $\,v_0  \,$  the initial value of v,  system (\ref{12}) can be given the form of the following integro differential equation

  \beq            \label{11}
     {\cal L}\,u\, \equiv \,  u_t -  \varepsilon  u_{xx} + au +b \int^t_0  e^{- \beta (t-\tau)}\, u(x,\tau) \, d\tau = \, F(\,x,\,t,\,u \,(x,t)\, ) \,    
 \eeq

\no  as soon as one puts: 

\beq     \label{14}
F(x,t,u)\, =\,\varphi (u) \, -\, v_0(x) \, e^{\,-\,\beta\,t\,}.
\eeq

\vs Equation  (\ref{11})  describes  the evolution of several  physical models  as motions of viscoelastic fluids or solids  \cite{bcf,dr1,r}; heat conduction at low temperature \cite{mps}, sound propagation in viscous gases \cite{l}.

\vs Moreover, it occurs also in superconductivity to describe the  Josephson tunnel effects in junctions. In this case the unknown  $\, u\, $ denotes the difference between the phases of the wave functions of the two superconductors and it results:

\beq  \label{15}
\varepsilon u_{xxt}\, - \, u_{tt} \, +\, u_{xx}-\, \alpha u_t = \,  \, \sin u \ + \gamma   
\eeq 

\vs \no where $\, \gamma \, $ is a constant  forcing term that is proportional to a bias current, and the  $ \varepsilon -term$  and the $\, \alpha -term $ account for  the dissipative normal electron current  flow along and across the junction, respectively \cite{bp}.

From (\ref{11}) one obtains  the equation (\ref{15})  as soon as one assumes

\beq   \label{16}
 a \,=\,  \alpha \, - \dfrac{1}{\varepsilon} \, \,\quad\,\, b = \,  - \, \dfrac{a}{\varepsilon}  \,\,\quad \displaystyle \, \beta \,= \dfrac{1}{\varepsilon}\,\,
  \eeq
  
  \no  and $ F  $ is such that

\beq   \label{17}
F(x,t,u)\,=\, -\, \int _0^t \, e^{\,-\,\frac{1}{\varepsilon}\,(t-\tau\,)}\,\,[\, \, \gamma\,+\, sen \, u (x, \tau)\, \,]\, \, d\tau. 
\eeq

\vs As (\ref{16}) show, in the superconductive case the constants $ \,a,\,  b\,$ could be negative too.

\vs In this paper Neumann and the Dirichlet problem for (1.3) are considered. By
means of the fundamental solution K0(x,t) of the operator L, which has already been determined
in [21], in the linear case, the explicit solutions are obtained. When the source term
F is non linear, an appropriate analysis of the integro differential equation implies
results on the existence and uniqueness of solutions. Besides, in both the linear case and the
non linear problem, a priori estimates of the solutions are achieved. The results are
applied  to (FHN) system.

\section{Statement of the problems and Laplace transform}
\setcounter{equation}{0}

If  $\, T\, $  is  an arbitrary positive constant and 

\[
\,   \Omega_T \, \equiv \{\,(x,t) : \, 0\,\leq \,x \,\leq L \,\,;  \ 0 < t \leq T \, \},
\]

\vs\no let ($ P_{N} $) the  following Neumann initial - boundary value  problem  related to equation (\ref{11}):

\beq   \label{21}
\left \{
   \begin{array}{lll}
   u_t -  \varepsilon  u_{xx} + au +b \int^t_0  e^{- \beta (t-\tau)}\, u(x,\tau) \, d\tau \,=\, F(x,t,u) \, & (x,t) \in \Omega_T \,  \\    

  \,u (x,0)\, = u_0(x)\, \,\,\, &
x\, \in [0,L], 
\\
 
  \, u_x(0,t)\,=\,\psi_1(t)  \qquad u_x(L,t)\,=\,\psi_2(t)  & 0<t\leq T.
   \end{array}
  \right.
\eeq

\vs\vs  \no In excitable systems this problem occurs when two-species reaction diffusion system is subjected to flux boundary condition \cite{m2}. The same conditions  are present in case of pacemakers \cite{ks}.  Neumann  conditions are applied also to study distributed FHN system \cite{ns}.

\vs\no Besides, if the Dirichlet conditions are  applied, the following initial boundary value problem $(P_D)$  holds: 

\beq   \label{21}
\left \{
   \begin{array}{lll}
{  \cal L}\,u\,  \,=\, F(x,t,u) \, & (x,t) \in \Omega_T \,  \\    
  \,u (x,0)\, = u_0(x)\, \,\,\, &
x\, \in [0,L], 
\\

  \, u(0,t)\,=\,g_1(t)  \qquad u(L,t)\,=\,g_2(t)  & 0<t\leq T.
   \end{array}
  \right.
\eeq

\no For example, in mathematical biology, those boundary  conditions occur when  the behavior  of a single dendrite has to determine and the voltage level has to be fixed.\cite{ks}

\vs When   $ F\,= f(x,t) $ is a linear function,  problems $ (P_{N}),(P_{D}) $  can  be solved by Laplace transform with respect to $ t. $

  If

 \beq   \label{22}
 \left \{
   \begin{array}{lll} 
\hat u (x,s) \, = \int_ 0^\infty \, e^{-st} \, u(x,t) \,dt \,& \hat f (x,s)   \, = \int_ 0^\infty \,\, e^{-st} \,\, f\,(x,t) \,dt \,,
\\ 
\\
\hat \psi_i \,\,(s) \, = \int_ 0^\infty \, e^{-st} \,\, \psi_i\,(t)\, \,dt \,\,&(i=1,2),
 \end{array}
  \right.
\eeq

\vs\no one deduces the following transform   $ ( \hat P_N) $ problem:

\beq   \label{23}
\left \{
   \begin{array}{lll}
  \hat u_{xx}  \,\,- \frac{\sigma^2}{\varepsilon} \,\,\hat u =\, -\,\frac{1}{\varepsilon} \,\,[\, \,\hat f(x,s) +u_0(x)\,\,]\\    
\\
  \,\hat u_x(0,s)\,=\, \hat \psi_1\,(s)\qquad \hat u_x(L,s)\,=\,\hat \psi_2\,(s),
   \end{array}
  \right.
\eeq

\vs  \no    where $ \,\,\,\ds \sigma^2 \ \,=\, s\, +\, a \, + \, \frac{b}{s+\beta}.\,\,\,$  Letting  $\displaystyle\,\tilde{\sigma}^2\,=\, \sigma^2/{\varepsilon,}\,\, $ and considering the following function

\beq\,  \label{24}
\displaystyle
\hat \theta_0 \,(\,y,\tilde\sigma)\,= \,\dfrac{\cosh\,[\, \tilde\sigma \,\,(L-y)\,]}{\,2\, \, {\varepsilon} \,\,\tilde \sigma\,\,\, \sinh\, (\,\tilde \sigma \,L\,)}\,\,=
\eeq

\[ =\,  \frac{1}{2 \,\, \sqrt\varepsilon \,\,\,\sigma  } \, \biggl\{\, e^{- \frac{y}{\sqrt \varepsilon} \,\,\sigma}+\, \sum_{n=1}^\infty \,\, \biggl[ \,e^{- \frac{2nL+y}{\sqrt \varepsilon} \,\,\sigma} \, +\, e^{- \frac{2nL-y}{\sqrt \varepsilon} \,\,\sigma}\,
\biggr] \, \biggr\},    \]

\vs \vs \no the   formal solutions $ \hat u(x,s)$ of  the problems $ ( \hat P_N) $  can be given the form:

\beq     \label{25}
\begin{split}
\hat u (x,s) = &\,\int _0^L \, [\,\hat \theta_0\,(\,|x-\xi|, \,s\,)\,+\,\,\,\hat \theta_0\,(\,|x+\xi|,\, s\,)\,] \, \,[\,u_0(\,\xi\,) \,+\,\hat f(\,\xi,s)\,]\,d\xi\, -
\\ 
  \,&  -\,\,\ 2 \,\,\varepsilon \, \,\hat \psi_1 \,(s) \,\, \hat  \theta_0 (x,s)\,+ \, 2 \,\, \varepsilon  \,\, \hat \psi_2 \, (s)\,\,\hat  \theta_0 \,(L-x,s\,).\, \,\,
\end{split}
\eeq 

\vs \no Analogously, for problem $ (\hat P_D) $, one obtains:  
  
\beq     \label{25}
\begin{split}
\hat u (x,s) = &\,\int _0^L \, [\,\hat \theta_0\,(\,x+\xi, \,s\,)\,-\,\,\,\hat \theta_0\,(\,|x-\xi|,\, s\,)\,] \, \,[\,u_0(\,\xi\,) \,+\,\hat f(\,\xi,s)\,]\,d\xi\, -
\\ 
  \, & -\,\,\ 2 \,\,\varepsilon \, \,\hat g_1 \,(s) \,\,   \dfrac{\partial}{\partial \,x}\,\, \hat\theta_0 (x,s)\,+ \, 2 \,\, \varepsilon  \,\, \hat \psi_2 \, (s)\,\,\dfrac{\partial}{\partial \,x}\,\,\hat  \theta_0 \,(L-x,s\,).\, \,\,
\end{split}
\eeq 
 
\vs \section{ Fundamental solution $K_0 (x,t)$ and theta function $ \theta_0(x,t) $}  

\setcounter{equation}{0}
\setcounter{figure}{0}

\setcounter{definition}{0}
\setcounter{notation}{0}
\setcounter{theorem}{0}
\setcounter{esempio}{0}
\setcounter{ex}{0}
\setcounter{proposition}{0}
\setcounter{criterio}{0}

\vs The fundamental solution $ K_0(x,t)  $ of the linear operator  $\, \cal L \,$ defined in (\ref{11}) has been already obtained in \cite{dr8} and  it results:

 \beq      \label {31}
 K_0(r,t)=  \frac{1}{2 \sqrt{\pi  \varepsilon } }\biggl[ \frac{ e^{- \frac{r^2 }{4 t}-a\,t}}{\sqrt t}-\,\sqrt b \int^t_0  \frac{e^{- \frac{r^2}{4 y}\,- ay}}{\sqrt{t-y}} \, e^{-\beta (\, t \,-\,y\,)}  J_1 (2 \sqrt{\,by\,(t-y)\,})\,\,dy \biggr],
\eeq 

\no where $\, r \, = |x| \, / \sqrt \varepsilon \, \, $ 
 and  $ J_n (z) \,$  is  the Bessel function of first kind with

\beq      \label{32}
\,{\cal L  }_t\,\,K_0\,\equiv \,\,\int_ 0^\infty e^{-st} \,\, K_0\,(r,t) \,\,dt \,\,=  \,
 \frac{e^{- \,r\,\sigma}}{2 \, \sqrt\varepsilon \,\sigma \,  } \qquad  \Re e  \,s > \,max(\,-\,a ,\,-\beta\,)\,.
\eeq

\no Indicating by  $\,\, \omega = min(a,\beta)\,\,$ and

\vs
\beq      \label{34} 
E(t) \,=\, \frac{e^{\,-\,\beta t}\,-\,e^{\,-\,at}}{a\,-\,\beta}\,\,>0\,,\qquad \beta _0 =\,\, \frac{1}{a}\, +\, \pi \sqrt b \, \, \displaystyle {\frac{a+\beta}{2(a\beta)^{3/2}}},
\eeq
\vs \no it results \cite{dr8}: 

\beq               \label{355}
|K_0| \, \leq \, \frac{e^{- \frac{r^2}{4 t}\,}}{2\,\sqrt{\pi \varepsilon t}} \,\, [ \, e^{\,-\,at}\, +\, b t \,E(t)\, ];  \quad \qquad \int_0^t\,d \tau\, \int_\Re |K_0(x-\xi,t)| \, \,d\xi \leq \,  \beta_0,
\eeq

\beq               \label{366}
\int_\Re\,\,|\,K_0(x-\xi,t\,)\,|\,\,d\xi\,\,\leq \, e^{\,-\,at}\, +\, \sqrt b\, \pi \,t \,  \, e^{\,-\,\omega \, t }. \,
\eeq

\vs \vs  In order to obtain the inverse formulae for (\ref{25}),          let apply (\ref{32}) to (\ref{24}). Then  one deduces the  following function similar to  {\em theta functions}:

\beq     \label{37}
\theta_0 (x,t) \,=\,  K_0(x,t) \ +\, \sum_{n=1}^\infty \,\, \ [\, K_0(x \,+2nL,\,t) \, + \, K_0 ( x-2nL, \,t)\,] \, =  
\eeq
\[\, =\sum_{n=-\infty }^\infty \,\, \ K_0(x \,+2nL,\,t). \,\]

 \vs As consequence, by (\ref{25}), the explicit solution of  the {\em linear} problem $\, (P_N )\,$ where $\, F\, =\, f(x,t) \,$ is :

\beq   \label{38}
 u(\, x,\,t\,)\, = \,\,\int^L_0 \, [\theta_0 \,(|x-\xi|,\, t)\,+ \theta_0 (x+\xi,\,t)\,]\, \,u_0(\xi)\,\, d\xi \,\,+ \,
\eeq

\[ - \,2 \, \varepsilon \,\int^t_0 \theta_0\, (x,\, t-\tau) \,\,\, \psi_1 (\tau )\,\,d\tau\,+\, 2\,\, \varepsilon \int^t_0 \theta_0\, (L-x,\, t-\tau) \,\,\, \psi_2 (\tau )\,\,d\tau\,\]

\[ +\, \,\int^t_0 d\tau\int^L_0 \, [\,\theta_0\, (|x-\xi|,\, t-\tau)+ \theta_0 (x+\xi,\,t-\tau )] \,\,\, f\,(\,\xi,\tau\,)\, \,\,d\xi.\]

\vs\, 
In an analogous way, a similar formula  for the problem $ (P_D) $ can be obtained:

\beq   \label{389}
 u(\, x,\,t\,)\, = \,\,\int^L_0 \, [\theta_0 \,(x+\xi,\, t)\,- \theta_0 (|x-\xi|,\,t)\,]\, \,u_0(\xi)\,\, d\xi \,\,+ \,
\eeq

\[ - \,2 \, \varepsilon \,\int^t_0  \dfrac{\partial}{\partial \,x}\,\,\theta_0\, (x,\, t-\tau) \,\,\, g_1 (\tau )\,\,d\tau\,+\, 2\,\, \varepsilon \int^t_0 \dfrac{\partial}{\partial \,x}\,\,\theta_0\, (L-x,\, t-\tau) \,\,\, g_2 (\tau )\,\,d\tau\,\]

\[ +\, \,\int^t_0 d\tau\int^L_0 \, [\,\theta_0\, (x+\xi,\, t-\tau)- \theta_0 (|x-\xi|,\,t-\tau )] \,\,\, f\,(\,\xi,\tau\,)\, \,\,d\xi.\]

\vs Owing to the basic properties of $ K_0(x,t), $ it is easy to deduce the following theorems:

\vs  \bthe
When the linear source $ \, f(x,t)\,  $ is continuous in $ \Omega_T\, $ and the initial boundary  data $ u_0(x),\,\, \psi_i\,(t)\, \,(i=1,2)\, \,   $ [ $ g_i\,\,(i=1,2)$]  are continuous, then  problem $ (P_N) $ [$ (P_D) $] admits a unique regular solution $ u(x,t)  $ given by (\ref{38})[(\ref{389})].
\hbox{}\hfill\rule{1.85mm}{2.82mm}
\ethe

\vs As consequence of the properties of fundamental solution $ K_0 (x,t),$ various estimates for $ u,\,\, u_t,\,\,u_x...  $ could be
obtained. 

As an example, let evaluate  the asymptotic properties of the terms caused by the initial datum $ u_0(x) $  and the source $ f(x,t).\,\, $ If

\[\,\,\,||\,u_0\,|| \,= \displaystyle \sup _{ 0\leq\,x\,\leq \,L\,}\, | \,u_0 \,(\,x\,) \,|, \,\,\qquad||\,f\,|| \,= \displaystyle \sup _{ \Omega_T\,}\, | \,f \,(\,x,\,t\,) \,|, \, \]

\vs \no it results:

\bthe
  When $ \psi_i\,=\,0 \, \,\,\,(i=1,2),\, $ $[g_i =0\,\,(i=1,2) ]$  the solution (\ref{38})[(\ref{389})] of $ (P_N) $,[ $ (P_D) $] for large $ t , $  verifies  the following estimate:

 \beq   \label{399}
|u(x,t)| \, \leq \,\, 2 \,\,\bigl[\,\,||\,f\,|| \,\, \beta_0\,+\, \,  \,\,||\,u_0\,|| \,\, ( 1\, +\, \sqrt b \,\pi \,t \, ) \,\,e^{- \omega \, t\,}\, \bigr]
\eeq

\vs \no  where $ \, \omega = \min \,(a, \beta )$ and  $ \beta_0\,$
is given by $\beta _0 =\,\, \frac{1}{a}\, +\, \pi \sqrt b \, \, \displaystyle {\frac{a+\beta}{2(a\beta)^{3/2}}}$

\ethe
Proof: Properties of $ K_0(x,t)$ imply that:

\vs 
\beq               \label{310}
\biggr| \,  \int_0^L\,\theta_0 (|x-\xi|,\,t)\,\,d\xi\,\biggl|\,\,\, \leq \,\sum_{n= -\infty }^\infty \, \, \int_0^L\,| K_0(|x-\xi +2nL|, \,t)| \,\,d\xi\,\ =
\eeq

\vs 
\[ =\,\sum_{n= -\infty }^\infty \, \, \int_{x+(2n-1)L}^{x+2nL}\,| K_0(y,\,t)| \,dy\,\,\,\,\leq  \,\, \,\int_\Re\,\,|K_0(y,t)|\,d y .\,\,\]

\vs \vs \no So,  applying properties   $ (\ref{355})_2   $ and (\ref{366})  to  (\ref{38}) [(\ref{389})], the  estimate (\ref{399}) follows. 
\hbox{}\hfill\rule{1.85mm}{2.82mm}

\vs
\section{The FitzHugh-Nagumo model. A priori estimates }

\setcounter{equation}{0}
\setcounter{figure}{0}

\setcounter{definition}{0}
\setcounter{notation}{0}
\setcounter{theorem}{0}
\setcounter{esempio}{0}
\setcounter{ex}{0}
\setcounter{proposition}{0}
\setcounter{criterio}{0}

\vs {\it Linear case - } If  the reaction kinetics of the model can be outlined by means of piecewise linear approximations, then one has:

\vspace {3mm}
\beq                 \label{40}
f(u)= \,\,\eta \, (\, u-a \,) \, \,-u\, \,\,\,\qquad\qquad\, \ (\,0\,<\,a\,< \,1\,)
\eeq

\vspace {3mm}\noindent where $ \,\eta \,$ denotes the unit- step function \cite {mk,t}. Like in   (\ref{13}), the linear approximation (\ref{40}) involve a linear term $\,\, -\,\,u.\,\,\,$ As consequence, denoting by $\,\,\bar \eta \,\, $  the constant that  holds zero or one, the FHN model  can be given the form:

\vspace{3mm}
  \beq                                                     \label{}
  \left \{
   \begin{array}{lll}
    {u_t } \,-  \varepsilon \, u_{xx} + \,u + \,v \,   = \, \bar \eta \,  
\\    \hspace{7cm}  (x,t)\in \Re\\
 v_t  \, + \beta \, v \, - b\, u\,=\,0 \,\,\, \,,
\\
      \end{array}
  \right.
 \eeq

\vs \no  and estimate (\ref{399}) can be applied.

\vs {\it Non linear case - }Consider now the non linear case  defined by (\ref{13}). By means of the previous results we are able to obtain integral equations for the two components  $ (u,v) \,$   in terms of the data.  All this implies, also in this case, a qualitative analysis of the solution  together with a priori estimates.

\vs \no At first let us observe that by $(\ref{12})_2 $ one has:

\beq      \label{41}
v\, =\,v_0 \, e^{\,-\,\beta\,t\,} \,+\, b\, \int_0^t\, e^{\,-\,\beta\,(\,t-\tau\,)}\,u(x,\tau) \, d\tau
\eeq

\vs \no and this formula, together with (\ref{14}) require the presence of  the following convolutions:

 \beq     \label {42}
 K_i( r,t) \, = \,\,\int^t_0 \,\,e^{-\,\beta \,(\,t-\tau)\,}\,K_{i-1}\,(x,\tau\,) \, d \tau\,\, \qquad ( i=1,2) 
 \eeq

\vs \vs \no which explicitly are given by \cite{dr8}:

\beq   \label{43}
K_i = \int_0^t\,\frac{e^{- \frac{x^2}{4\varepsilon y}- a\,y\,-\,\beta(t-y)}}{2\,\sqrt{\,\pi\,\varepsilon \,y}} \, \biggl(\sqrt{\frac{t-y}{b\,y}} \biggr)^{i-1}\, J_{i-1 } (2 \sqrt{\,b\,y\,(t-y)\,})\,dy \, \,\,\,\, \,\, (i=1,2).
\eeq

\vs\vs \no As consequence, together with $ \theta_0 $ defined by (\ref{37}), the other two $ \theta  $ functions

\beq  \label{44}
 \theta_i (x,t)\,=\,  \sum_{n=-\infty }^\infty \,\, \ K_i(x \,+2nL,\,t) \, \qquad (i=1,2)
 \eeq

\vs \no must be considered. 

\vs  To allow a plainer reading let's set 
 
\beq      \label{45}
 G_i(x,\xi, t) \, = \, \theta_i\,(\,|x-\xi|, \,t\,)\,+\,\,\,\theta _i \,(\,x+\xi,\,t\,)\,\qquad (i=0,1,2)
\eeq

\vs \vs \no In this manner, owing to  (\ref{38}) one has:

\beq   \label{46}
 u(\, x,\,t\,)\, = \,\,\int^L_0 \, [\,G_0(x,\xi,\, t)\,\, \,u_0(\xi)\,\,- \,G_1(x,\xi,\, t)\, \,v_0(\xi)\,\,] d\xi \,+\,\eeq
\[ - \,2 \, \varepsilon \,\int^t_0 \theta_0\, (x,\, t-\tau) \,\,\, \psi_1 (\tau )\,\,d\tau\,+\, 2\,\, \varepsilon \int^t_0 \theta_0\, (L-x,\, t-\tau) \,\,\, \psi_2 (\tau )\,\,d\tau\,\]

\[ +\, \,\int^t_0 d\tau\int^L_0 \, \,G_0(x,\xi,\, t-\tau) \,\,\,  \varphi\,[\,\xi,\tau,\,u(\xi, \tau)]\,  \,\,d\xi\,.\,\]

\vs\,\vs  As for the  $ v $ component, by  (\ref{41}) one deduces:

 \beq      \label{47}
v(x,t) \, =\,\, v_0 \, e^{\,-\,\beta\,t\,}\,\,+\, b\, \,\,\int^L_0 \, [G _1\,(x,\xi,\, t)\,\, \,u_0(\xi)\,\,- G _2\,(x,\xi,\, t)\,\,  v_0(\xi)\,] d\xi \,\,+ \,
\eeq 
\[ - \,2 \,b\,\, \varepsilon \,\int^t_0 \theta_1\, (x,\, t-\tau) \,\,\, \psi_1 (\tau )\,\,d\tau\,+\, 2\,\,b\,\, \varepsilon \int^t_0 \theta_1\, (L-x,\, t-\tau) \,\,\, \psi_2 (\tau )\,\,d\tau\,\]

\vs
\[ +\,b  \,\int^t_0 d\tau\int^L_0 \, \,G_1\, (x,\xi,\, t-\tau)\,\,\, \varphi\,[\,\xi,\tau,\, u(\xi, \,\tau)\,]\,d\xi.\,\, \]

\vs\vs\vs  Let us observe that the kernels $ K_1(x,t)  $ and $ K_2(x,t)  $ have the same  properties of $ K_0(x,t).  $ In fact \cite{dr8}:

\vs \bthe

For all the positive constants $ a, \,b\, \varepsilon, \beta $  it results:

\beq   \label{48}
\int_\Re |K_1| \, \ d\xi \leq \, \,E(t);\,  \qquad  \int _0^t\\d\tau \,\int_\Re |K_1| \, \ d\xi \leq \, \beta_1\,
\eeq

\beq   \label{49}
\int_\Re \left|K_2 (x-\xi,t)\right| \, d\xi \, \leq \, \int_0^t \, \, e^{\,-\,ay\, - \beta(t-y)\,}\,(\,t-y\,) \,dy \,\leq \, t\, E(t)
\eeq

\vs \vs \no where $ E(t) $ is defined in $(\ref{34})_1$ and $ \beta_1\,=\, ({a\,\beta})^{\,-1}.\,$ \hbox{}\hfill\rule{1.85mm}{2.82mm}
\ethe

\vs\vs  Now, let $\,\, ||\,z\,|| \,= \displaystyle \sup _{ \Omega_T\,}\, | \, z \,(\,x,\,t) \,|, \,\, $ and let $ \,{\cal B}_ T \,  $ denote the Banach space 

\beq   \label{35}
  \,{\cal B}_ T \, \equiv \, \{\, z\,(\,x,t\,) : \, z\, \in  C \,(\Omega_T),  \, \,\,   ||z|| \, < \infty \ \}.
\eeq

\vs\vs   By means of standard methods related to integral equations  and owing to basic properties of $ K_i,\,\, G_i \,\, (i=0,1,2)$  and $ \varphi(u),\,\, $it is easy  to prove that the  mapping defined by (\ref{46})[(\ref{47})] ia a contraction  of $ {\cal B}_ T $ in $ {\cal B}_ T  $ and so it admits an unique fixed point   $ u(x,t)  \, \in {\cal B}_ T $  \cite{c,dmm}.  Hence

\vs
\bthe
When the initial data $ (u_0, v_0 )$ are continuous functions, then the Neumann [Dirichlet] problem related to the non linear (FHN) system (\ref{12}),(\ref{13}) has a unique solution in the space of  solutions which are regular in $ \Omega_T $. 
 \hbox{}\hfill\rule{1.85mm}{2.82mm}
 \ethe

\vs \vs Continuous dependence for the solution of  $ (P_N) $ [$ (P_D) $] is an obvious consequence of the previous estimates.

As for asymptotic properties, it is well known  that the  (FHN) system admits arbitrary large invariant rectangles
$ \Sigma $ containing $(0,0)$ so that the solution $(u,v)$, for all times $t > 0$, lies in the interior of $ \Sigma $
 when the initial data $(u_o,v_o)$ belong to $ \Sigma $. \cite{s}  For this,  considering,   for example, the case  $ \psi_1\,=\,\psi_2\,=\,0\, $ and letting

\[
||\,\varphi\,|| \,= \displaystyle \sup _{ \Omega_T\,}\, | \,\varphi \,(\,x,\,t,\,u) \,|, \, \]

\vs \no then by means  of (\ref{46}), (\ref{47}) and  owing to the estimates (\ref{35}), (\ref{36}), (\ref{48}), (\ref{49}), the following theorem can be stated:

\bthe
For regular solution $ (u,v) $ of the (FHN) model, when  $\psi_1\,\,= \psi_2 \,= \,0,\, \, $  the following estimates hold:

\vs 
\beq            \label{510}
\left\{ 
 \begin{array}{lll}                                                   
 \left| u \, \right| \, \leq  2\,[\,\left\| u_0 \right\| \, (1+\pi \sqrt b \, t ) \, e^ {\,-\omega\,t\,}\,+\,\left\| v_0 \right\|\,E(t) \, +\, \beta_0 \,\left\| \varphi \right\|\,] 
   \\
\\
\left| v \, \right| \, \leq  \left\| v_0 \right\|\, e^ {\,-\,\beta\,t\,}\,+\,2\,[\,b\,(\,\left\| u_0 \right\|\,+\, t\, \left\| v_0 \right\|\,) \, E(t) \, + \, b\, \beta_1\, \left\| \varphi \right\| \,]
\\ 
   \end{array}
  \right.
 \eeq
\hbox{}\hfill\rule{1.85mm}{2.82mm}
\ethe

\vs \no Therefore, when $ t $ is large, {\em the effect due to the initial disturbances $\, (\,u_0, v_0\,) \, $ is exponentially  vanishing  while   the effect of the non linear source is bounded for all $ t. $}

\vs \no Asymptotic effects of boundary perturbations can be found in \cite{dr13}. 

\vs \no All   the previous results  can be applied to the boundary  Dirichlet, too.

\vs \vs\no 
{\bf Acknowledgments}
This  work has been performed under the auspices of Programma F.A.R.O. (Finanziamenti per l' Avvio di  Ricerche
Originali, III tornata) ``Controllo e stabilita' di processi diffusivi nell'ambiente'', Polo delle Scienze e Tecnologie, Universita' degli Studi di Napoli Federico II  (2012).

\end{document}